\documentclass[aps,superscriptaddress,twocolumn,showpacs]{revtex4}
\usepackage{graphicx}

\begin{document}

\title{Surface location of sodium atoms attached to $^3$He nanodroplets}

\author{F. Stienkemeier}
\affiliation{Fakult\"at f\"ur Physik, Universit\"at Bielefeld, D-33615
Bielefeld, Germany}

\author{O. B\"unermann}
\affiliation{Fakult\"at f\"ur Physik, Universit\"at Bielefeld, D-33615
Bielefeld, Germany}

\author{R. Mayol}
\affiliation{Departament E.C.M., Facultat de F\'{\i}sica, Universitat
de Barcelona, E-08028, Spain}

\author{F. Ancilotto}
\affiliation{INFM (Udr Padova and DEMOCRITOS National Simulation
Center, Trieste, Italy) and Dipartimento di Fisica ``G. Galilei'',
Universit\`a di Padova, via Marzolo 8, I-35131 Padova, Italy}

\author{M. Barranco}
\affiliation{Departament E.C.M., Facultat de F\'{\i}sica, Universitat
de Barcelona, E-08028, Spain}

\author{M. Pi}
\affiliation{Departament E.C.M., Facultat de F\'{\i}sica, Universitat
de Barcelona, E-08028, Spain}

\begin{abstract}

We have experimentally studied the electronic  $3p\leftarrow 3s$
excitation of Na atoms attached to $^3$He droplets by means of
laser-induced fluorescence as well as beam depletion spectroscopy. From
the similarities of the spectra (width/shift of absorption lines) with
these of Na on $^4$He droplets, we conclude that sodium atoms reside in
a ``dimple'' on the droplet surface and that superfluid-related effects
are negligible. The experimental results are supported by Density
Functional calculations at zero temperature, which confirm the surface
location of sodium on $^3$He droplets, and provide a detailed
description of the ``dimple'' structure. The calculated shift of the
excitation spectra for the two isotopes is in good agreement with the
experimental data.


\end{abstract}

\date{\today}
\maketitle

Detection of laser-induced fluorescence (LIF) and beam depletion (BD)
signals upon laser excitation provides a sensitive spectroscopic
technique to investigate electronic transitions of chromophores
attached to $^4$He nanodroplets \cite{stienke2}. While most of atomic
and molecular dopants migrate to the center of the droplet, alkali
atoms (and alkaline earth atoms to some extent \cite{Sti:1997b}) have
been found to reside on the surface of $^4$He droplets, as evidenced by
the much narrower and less shifted spectra when compared to those found
in bulk liquid $^4$He \cite{scoles,ernst1,stienke3,Ernst:2001a}. This
result has been confirmed by  Density Functional (DF) \cite{anci1} and
Path Integral Monte Carlo (PIMC) \cite{nakayama} calculations, which
predict surface binding energies of a few Kelvin, in agreement with the
measurements of detachment energy thresholds using the free atomic
emissions \cite{KKL}. The surface of liquid $^4$He is only slightly
perturbed by the presence of the impurity, which produces a ``dimple''
on the underlying liquid. The study of these states can thus provide
useful information on surface properties of He nanodroplets
complementary to that supplied by molecular-beam scattering experiments
\cite{Dal98,har01}.

Although the largest amount of work has been devoted to the study of
pure and doped $^4$He nanodroplets -see \cite{toennies,kwon} and
Refs.~therein-, the only neutral Fermi systems  capable of being
observed as bulk liquid and droplets are made of $^3$He atoms, and for
this reason they have  also attracted the interest of experimentalists
and theoreticians \cite{harms,har01,panda,TF,Bar97a,Wei92,gar98,gua00}. We recall
that while $^4$He droplets, which are detected at an experimental
temperature ($T$) of $\sim$ 0.38\, K, are superfluid, these containing
only $^3$He atoms, even though detected at a lower $T$ of $\sim$
0.15\,K, do not exhibit superfluidity \cite{grebenev}.

The behavior of molecules in He clusters is especially appealing. In
particular, probes at the surface of the droplets are desirable because
they allow to investigate the liquid--vacuum interface as well as
droplet surface excitations. The latter are of interest in the
comparison of the superfluid vs. normal fluid behavior, particularly
because the Bose--Einstein condensate fraction has been calculated to
approach 100\% on the surface of $^4$He \cite{Griffin:1996}. Small
$^3$He drops are difficult to detect since, as a consequence of the
large zero-point motion, a minimum number of atoms is needed to produce
a selfbound drop \cite{panda,TF,Bar97a}. Microscopic calculations of $^3$He
droplets are scarce, and only concern the ground state (GS) structure
\cite{panda,gua00}. GS properties and collective excitations
of $^3$He droplets doped with some inert atoms and molecular impurities
have been addressed within the Finite Range Density Functional (FRDF)
theory \cite{gar98}, that has proven to be a valuable alternative to
Monte Carlo methods, which are notoriously difficult to apply to Fermi
systems. Indeed, a quite accurate description of the properties of
inhomogeneous liquid $^4$He at $T=0$ has been obtained within  DF
theory \cite{prica}, and a similar approach has followed for $^3$He
(see \cite{gar98,her02} and Refs.~therein).

The experiments we report have been performed in a helium droplet
machine used earlier for LIF and BD studies, and is described elsewhere
\cite{Sti:1997b}. Briefly, helium gas is expanded under supersonic
conditions from a cold nozzle forming a beam of droplets traveling
freely under high vacuum conditions. The droplets are doped downstream
employing the pick-up technique: in a heated scattering cell, bulk
sodium is evaporated in such a way that, on average, a single metal
atom is carried by each droplet. LIF absorption spectra of doped
droplets are recorded upon electronic excitation using a continuous
wave ring-dye laser and detection in a photo multiplier tube (PMT).
Since electronic excitation of alkali-doped helium droplets is
eventually followed by desorption of the chromophore, BD spectra can be
registered by a Langmuir-Taylor surface ionization detector
\cite{Sti:2000b}. Phase-sensitive detection with respect to the chopped
laser or droplet beam was used. For that reason the BD signal
(cf.~Fig.~\ref{exp_spectra}), i.e.~a decrease in intensity, is directly
recorded as a positive yield. For these experiments, a new droplet
source was built to provide the necessary lower nozzle temperatures to
condense $^3$He droplets. Expanding  $P_0=$ 20\,bar of helium gas
through a nozzle 5\,$\mu$m in diameter, we now can establish
temperatures down to 7.5\,K using a two-stage closed cycle refrigerator
(Sumitomo Heavy Industries, Model: RDK-408D). In this way, without
needing any liquid helium or nitrogen for pre-cooling or cold shields,
stable beam conditions can be utilized over several days.

\begin{figure}
\resizebox{1.0\columnwidth}{!}{\includegraphics{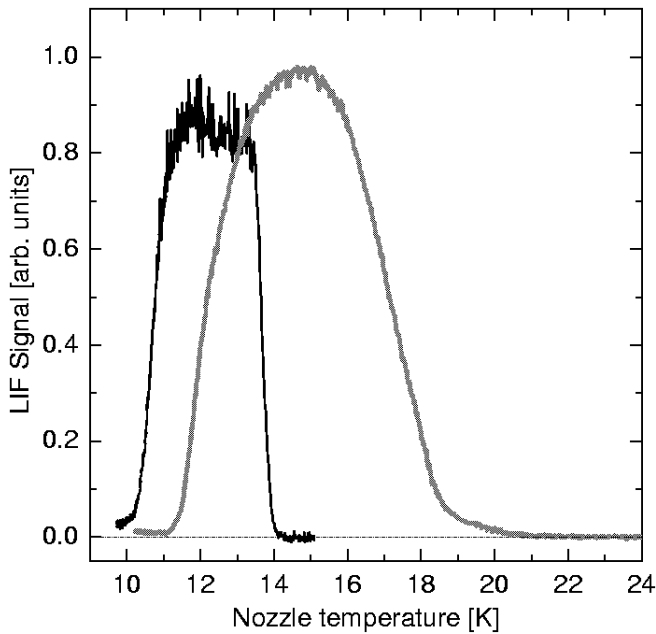}}
\caption{Laser-induced fluorescence signal as a function of nozzle
temperature forming $^3$He droplets (black) in comparison to $^4$He
droplets (grey). In both runs a stagnation pressure of 20\,bar was
used; nozzle diameter was 5\,$\mu$m. Normalization is such that the
plot gives the correct relative intensities.} \label{sourceconditions}
\end{figure}

\begin{figure}
\resizebox{1.0\columnwidth}{!}{\includegraphics{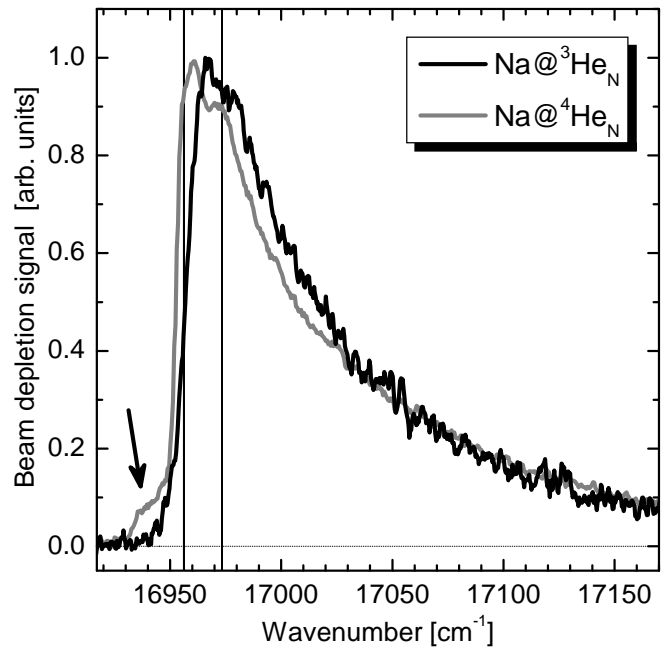}}
\caption{Beam depletion spectra of Na atoms attached to $^3$He/$^4$He
nanodroplets. The vertical lines indicate the positions of the two
components of the Na gas-phase  $3p\leftarrow 3s$
transition.}\label{exp_spectra}
\end{figure}

Fig.~\ref{sourceconditions} compares the number of Na-doped $^3$He vs.~$^4$He
droplets expanding 20\,bar helium as a function of the nozzle temperature
$T_0$. The absolute number densities in the maxima of both distributions are
quite similar. The formation of droplets from a supersonic expansion is well
known and thoroughly discussed in the literature: Besides the determination of
absolute cluster sizes \cite{Toe:unpublished,har01}, the size dependence as a
function of source conditions has already been studied \cite{Sch90,ernst1}. The
low temperature cutoff appears at source conditions where the isentrope of the
expansion hits the helium critical point \cite{Har97}; the disappearance at
high temperatures just means that the droplets are getting to small to carry
the dopant. Here, the critical size is determined by the thermal energy during
pick-up which leads to evaporation of helium atoms and a destruction of small
droplets. For the spectroscopic measurements presented in the following, we set
$T_0 = 11$\,K for $^3$He, and $T_0 = 15$\,K for $^4$He. These conditions are
expected to result in comparable mean cluster sizes around 5000 atoms per
droplet \cite{Toe:unpublished,har01}. As far as our results are concerned,
Fig.~\ref{sourceconditions} demonstrates that we see the correct formation as
well as droplet sizes of the different isotopes and that we use suitable source
conditions to guarantee comparable sizes.

In Fig.~\ref{exp_spectra} the absorption
spectrum of Na atoms attached to $^3$He nanodroplets is shown in
comparison to Na-doped $^4$He droplets. We present here the BD spectra
because they do not contain the strong fluorescence background lines of
free Na atoms which cover the crucial steep increase of the droplet
spectrum. Moreover, LIF does not always represent the total absorption
spectrum because it relies on the emission of a photon in the spectral
range of the PMT. Hence, absorption processes followed by either
radiationless decay or emission of photons in the infrared spectral
region are suppressed. The latter has been observed in LIF spectra
where alkali--helium exciplexes form upon excitation of alkali atoms on
the surface of $^4$He droplets \cite{stienke3,Lehmann:2000c}. In our
experiment Na$^3$He exciplexes are formed in the same way as their
Na$^4$He counterparts. This became immediately obvious because we were
able to discriminate the corresponding red-shifted emission
intensities. However, as far as the measured absorption spectra of
Na$@^3$He$_N$ are concerned, the LIF data are very well in accord with
the BD absorption.

The outcome of the spectrum of Na attached to $^3$He nanodroplets is
very similar to the spectrum on $^4$He droplets. The asymmetrically
broadened line is almost unshifted with respect to the gas-phase
absorption. This absence of a shift immediately confirms the surface
location because atoms embedded in bulk superfluid helium are known to
evolve large blue-shifts of the order of a couple of hundreds of
wavenumbers and much more broadened absorption lines
\cite{Takahashi:1993}. A blue shift is a consequence of the repulsion
of the helium environment against the spatially enlarged electronic
distribution of the excited state (``bubble effect''). The interaction
towards the $^3$He droplets appears to be slightly enhanced, evidenced
by the small extra blue shift of the spectrum compared to the $^4$He
spectrum. In a simple picture this means that more helium atoms are
contributing or, in other words, a more prominent ``dimple'' interacts with
the chromophore. The upper halves of the spectra are almost identical,
when shifting the $^3$He spectrum by $7.5\pm 1$\,cm$^{-1}$ to lower
frequencies. The influence of different droplet sizes does not affect
this observation: A dependence of the $^4$He spectra varying the
droplet size is already shown in \cite{scoles}. Probing smaller
droplets narrows the spectrum but does not shift the observed doublet
structure. The shift of the $^3$He spectrum would be even more
conspicuous when comparing smaller droplets. Unfortunately, the
corresponding small $^3$He droplet sizes are experimentally not
accessible because within the plateau region of the droplet yield shown
in Fig.~\ref{sourceconditions} the droplet sizes are almost unchanged
\cite{har01}. The increase in width (FWHM) of the $^3$He spectrum is
only 7\%. Taking into account the just mentioned droplet size
dependence of the width, this difference might not even be significant.
Regarding the substructure of the line, the only notable exception is
the absence of the red-shifted shoulder, which is observed in the case
of $^4$He and marked with an arrow in Fig.~\ref{exp_spectra}. This
feature, which is even more pronounced in the absorption of Li-doped
$^4$He droplets \cite{scoles} has not been interpreted yet.  The shift
with respect to the maximum of the absorption line is
$\approx\,20\,$cm$^{-1}$, too high in energy to be attributed to an
excited compressional or surface mode of the droplet at 0.38 K
\cite{gar98,Chi95}.

\begin{figure} 
\resizebox{\columnwidth}{!}{\includegraphics{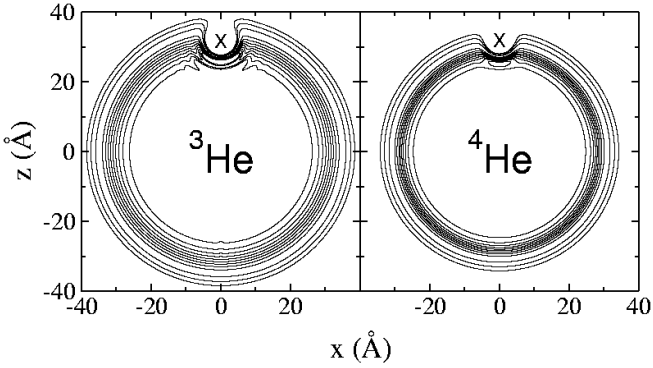}}
\caption{Equidensity lines in the $x-z$ plane showing the stable state
of a Na atom (cross) on a He$_{2000}$ droplet. The 9 inner lines
correspond to densities $0.9 \rho_0$ to $0.1 \rho_0$, and the 3 outer
lines to $10^{-2} \rho_0$, $10^{-3} \rho_0$, and  $10^{-4} \rho_0$
($\rho_0=0.0163$ \AA$^{-3}$ for $^3$He, and 0.0218 \AA$^{-3}$ for
$^4$He). } \label{fig3}
\end{figure}

FRDF calculations at $T=0$ confirm the picture emerging from the
measurements, i.e.~the surface location of Na on $^3$He nanodroplets
causing a more pronounced ``dimple''  than in $^4$He droplets. We have
investigated the stable configurations of a sodium atom on both $^3$He
and $^4$He clusters of different sizes. The FRDF's used for $^3$He and
$^4$He are described in \cite{bar97,may01}. The large number of $^3$He
atoms we are considering allows to use the extended Thomas-Fermi
approximation \cite{TF}. The minimization of the energy DF's with
respect to density variations, subject to the constraint of a given
number of He atoms $N$, leads to Euler-Lagrange equations whose
solution give the equilibrium particle densities $\rho ({\bf r})$.
These equations have been solved as indicated in \cite{Bar03}. The
presence of the foreign impurity is modeled by a suitable potential
obtained by folding the helium density with a Na-He pair potential.
We have used the potential proposed by Patil \cite{patil}
to describe the impurity-He interactions. Potential energy curves
describing the He-alkali interaction have been calculated recently by
ab-initio methods \cite{nakayama}, and found to agree very well with
the Patil potential.

\begin{figure} 
\resizebox{1.0\columnwidth}{!}{\includegraphics{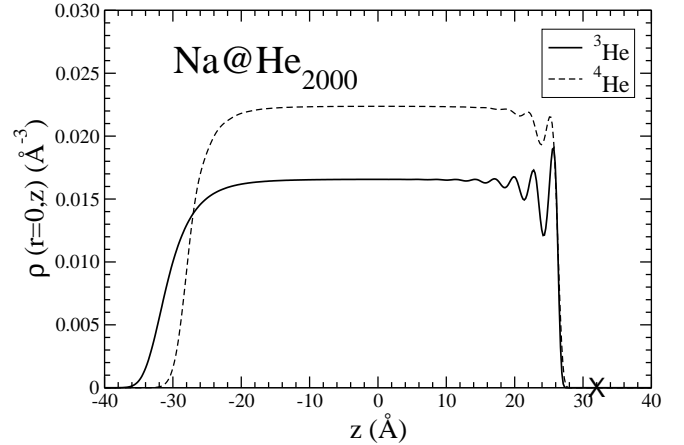}}
\caption{Density profiles along a line connecting the impurity to the
center of the cluster showing the equilibrium configuration of a Na
atom (cross) on He$_{2000}$ nanodroplets.} \label{fig4}
\end{figure}

Fig.~\ref{fig3} shows the equilibrium configuration for a Na atom
adsorbed onto He$_{2000}$ clusters. For a given $N$, the size of the
$^3$He$_N$ droplet is larger than that of the $^4$He$_N$ droplet -an
obvious consequence of the smaller $^3$He saturation density-.
Comparison with the stable state on the $^4$He$_{2000}$ cluster  shows
that, in agreement with the experimental findings presented before, the
``dimple'' structure is  more pronounced in the case of $^3$He, and that
the Na impurity lies {\it inside} the surface region for $^3$He and
{\it outside} the surface region for $^4$He
(we recall that the surface region is usually defined as that comprised 
between the
radii at which $\rho=0.1 \rho_0$ and $\rho=0.9 \rho_0$, where $\rho_0$
is the He saturation density \cite{Dal98,har01,TF}).
We attribute this to the
lower surface tension of $^3$He (0.113 K/{\rm \AA}$^2$) as compared to
that of $^4$He (0.274 K/{\rm \AA}$^2$), which also makes the surface
thickness of bulk liquid and droplets larger for $^3$He than for $^4$He
\cite{Dal98,har01}. The Na-droplet equilibrium distance, here defined
as the `radial' distance  between the impurity and the point where the
density of the {\it pure} drop would be $\rho \sim \rho_0/2$,
is $R \sim 1.1$ \AA\, for
$^3$He, and $R \sim 3.6$ \AA\, for $^4$He. For the larger droplets we
have studied, $R$ is nearly $N$-independent. A related quantity is the
deformation of the surface upon Na adsorption, which can be
characterized \cite{anci1} by the ``dimple'' depth, $\xi$, defined as the
difference between the position of the dividing surface at $\rho \sim
\rho_0/2$, with and without impurity, respectively. We find $\xi \sim  4.5$
\AA \,for $^3$He, and $\xi \sim 2.1$ \AA $\,$ for $^4$He. Fig.~\ref{fig4}
shows the  density profiles for Na$@$He$_{2000}$. Note the more diffuse
liquid-vacuum interface for the $^3$He droplet far from the impurity,
and the occurrence of more marked density oscillations (with respect to
the $^4$He case) where the softer $^3$He surface is compressed by the
adsorbed Na atom. Our calculations thus yield the detailed structure of
the liquid around the impurity, which is an essential ingredient for
any line-shift calculation of the main electronic transitions in the
adatom \cite{scoles}
and for the
understanding of dynamical processes which already have been observed
in time-dependent experiments \cite{Sti:unpublisheda}.

We have obtained the shift between the $^3$He and $^4$He spectra in
Fig.~\ref{exp_spectra} within the Frank-Condon approximation, i.e.
assuming that the ``dimple'' shape does not change during the Na excitation
The shift is calculated within the model given in \cite{Kan94},
evaluating Eqs.~5 and 6 therein, both for $^3$He and $^4$He. We used the
excited state A $^2\Pi$ and B $^2\Sigma$ potentials of \cite{nakayama}
because their Na-He GS potential is very similar to the Patil
potential we have used to obtain the equilibrium configurations.
For our largest droplet ($N=2000$) we find that the
$^3$He spectrum is blue-shifted with respect to the $^4$He one by
6.4\,cm$^{-1}$, in good agreement with the experimental value of
$7.5\pm 1$\,cm$^{-1}$ as extracted from Fig.~\ref{exp_spectra}.

\begin{figure} 
\resizebox{1.0\columnwidth}{!}{\includegraphics{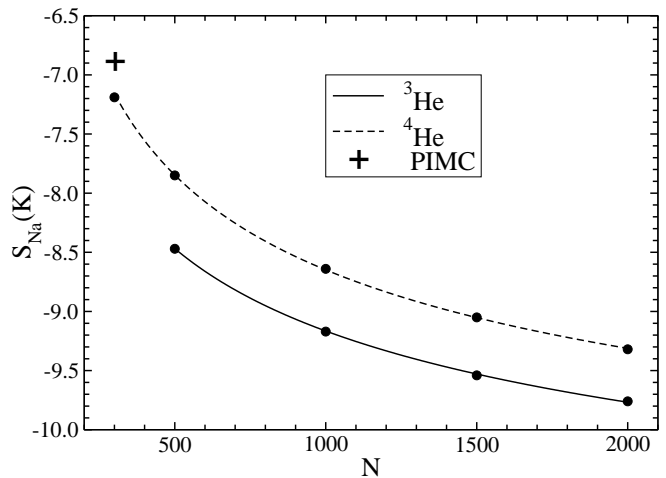}} \caption{ Na
solvation energy as a function of $N$. The cross is the PIMC result of
\cite{nakayama}. 
} \label{fig5}
\end{figure}

Solvation energies $ S_{Na}=E({\rm Na}@{\rm He}_N)-E({\rm He}_N) $ are
shown in Fig.~\ref{fig5} as a function of droplet sizes. More negative
values of $S_{Na}$ indicate increased binding of the Na atom to the
droplet surface. 
The lines are a fit of the form 
\begin{equation}
S_{Na}(N)= S_0+
\frac{S_1}{N^{1/3}} + \frac{S_2}{N^{2/3}}
\end{equation}
with $S_0=-12.1 (-12.5)$, $S_1=-28.3
(-31.6)$, and $S_2=37.3 (37.6)$ K for $^4$He($^3$He). 
To compare with ``exact'' PIMC result \cite{nakayama},
we have calculated Na$@^4$He$_{300}$. Part of the small difference
between the two values has to be attributed to our neglecting of the
Na zero-point energy. The value $S_{Na} \sim -12$ K has been
obtained within FRDF theory for Na adsorbed on the {\it planar} surface
of $^4$He \cite{anci1}, which corresponds to the $N=\infty$ limit
in this figure.

Our results show that Na adsorption on $^3$He droplets occurs in
very much the same way as in the case of $^4$He, i.e., the adatom is
located on the surface, though in a slightly more pronounced ``dimple''.
The similarities in the experimental spectra are certainly remarkable
for two apparently very different fluids, one normal and the other
superfluid, and clearly indicate that superfluidity does not play any
substantial role in the processes described here. This is likely a
consequence of the very fast time scale characterizing the Na
electronic excitation compared to that required by the He fluid to
readjust. The excitation occurs in a ``frozen'' environment and the
only significant difference between $^3$He and $^4$He is due to the
different structure of the``dimple'', which accounts for the small
shift in their spectra observed in the experiments and
found in our calculations as well. 

We thank Flavio Toigo for useful comments. This work has been supported
by grants MIUR-COFIN 2001 (Italy), BFM2002-01868 from DGI (Spain), and
2001SGR-00064 from Generalitat of Catalunya as well as the DFG
(Germany).

\end{document}